\documentclass[letterpaper]{JHEP3}

\newcommand{\reals}{\mathbb R}

\newtheorem{remark}{Remark} 

\title{A scalar invariant and 
the local geometry of a class of static spacetimes} 
\author{Manash 
Mukherjee\\ Department of Mathematics, University of California, 
Davis, CA 95616\\ E-mail: \email{manash@math.ucdavis.edu}}
\author{F. Paul Esposito\\ Department of Physics, 
University of Cincinnati, Cincinnati, OH 45221\\ E-mail: 
\email{esposito@physics.uc.edu}}
\author{L. C. R. Wijewardhana\\ 
Department of Physics, University of Cincinnati, Cincinnati, OH
45221 and Institute of Fundamental Studies, Kandy, 
Sri Lanka \\ E-mail: \email{rohana@physics.uc.edu}}

\abstract{The scalar invariant, $I\equiv 
R_{\mu\nu\rho\sigma;\delta}~R^{\mu\nu\rho\sigma;\delta}$, constructed 
from the covariant derivative of the curvature tensor is used to probe 
the local geometry of static spacetimes which are also Einstein 
spaces.  We obtain an explicit form of this invariant, exploiting the 
local warp-product structure of a 4-dimensional static spacetime, 
$~^{(3)}\Sigma \times_{f} \reals$, where $^{(3)}\Sigma $ is the 
Riemannian hypersurface orthogonal to a timelike Killing vector field 
with norm given by a positive function, $f :~^{(3)} 
\Sigma\longrightarrow \reals$.  For a static spacetime which is an 
Einstein space, it is shown that the locally measurable scalar, $I$, 
contains a term which vanishes if and only if $^{(3)}\Sigma$ is 
conformally flat; also, the vanishing of this term implies (a) 
$~^{(3)}\Sigma$ is locally foliated by level surfaces of $f$, 
$^{(2)}S$, which are totally umbilic spaces of constant curvature, and 
(b) $^{(3)}\Sigma$ is locally a warp-product space, $\reals\times$ 
$_{r(f)}$ $^{(2)}S$, for some function $r(f)$.  Futhermore, if 
$^{(3)}\Sigma$ is conformally flat it follows that every non-trivial 
static solution of the vacuum Einstein equation with a cosmological 
constant, is either Nariai-type or Kottler-type - the classes of 
spacetimes relevant to quantum aspects of gravity.}


\preprint{}
\begin {document}

\section{\sf Introduction}\label{sec1}

\par It is well known from the work of Cartan \cite{C} and Thomas 
\cite{T} that the local geometry of a semi-Riemannian manifold is 
uniquely determined up to isometries by the curvature tensor and its 
covariant derivatives up to a finite order.  This result, as suggested 
by Brans \cite{B} and further developed by Karlhede \cite{K1}, may be 
used to provide a coordinate independent characterization of 
gravitational fields in general relativity.  Based on this idea, 
Karlhede et al \cite{K2} have investigated the simplest scalar 
invariant, $I\equiv R_{\mu\nu\rho\sigma;\delta}~R^{\mu\nu\rho\sigma; 
\delta}$, constructed from the first covariant derivative of the 
curvature tensor, and found that the behavior of this locally 
measurable scalar reveals the effect of passage through the event 
horizon of the Schwarzschild spacetime - a static solution to the 
vacuum Einstein equation, $R_{\mu\nu} = 0$.  For any static 
spacetime, however, we observe that $I$ contains a term 
$\displaystyle{I_B}$ which is the ``square'' of the rank-$3$ tensor 
$\displaystyle{B\equiv\nabla_{V} R (V){\big |}_{\Sigma}}$ on the 
hypersurface $\Sigma$ orthogonal to the timelike Killing vector field 
$V$.  In particular, $\displaystyle{I_B}$ vanishes for the 
Schwarzschild spacetime where $\Sigma$ is known to be conformally flat.  
Interestingly, we find that the tensor $\displaystyle{\nabla_{V} R 
(V){\big |}_{\Sigma}}$ is proportional to the Cotton tensor \cite{E} 
on the spacelike hypersurface $\Sigma$, for {\it every} static solution to 
$R_{\mu\nu} = \Lambda g_{\mu\nu}$ - the vacuum Einstein equation with 
a cosmological constant $\Lambda$.  It is well-known that a 
semi-Riemannian $3$-manifold is conformally flat if and only if the 
Cotton tensor vanishes.  Hence, for all static solutions to 
$R_{\mu\nu} = \Lambda g_{\mu\nu}$, $\Sigma$ is conformally flat
if and only if $\displaystyle{I_B} = 0$. The principal goal of this paper 
is to determine all non-trivial {\it static spacetimes which are also
Einstein spaces} (henceforth, referred to as ``{\sf STE}'' spacetimes),
satisfying  $\displaystyle{I_B} = 0$. 

\par In fact, $\displaystyle{I_B}$ vanishes for a rich variety of {\sf 
STE} spacetimes (that is, static solutions to $R_{\mu\nu} = \Lambda 
g_{\mu\nu}$): ({\it a}) with $\Lambda = 0$, the Levi-Civita solutions 
in Class A \cite{Eh} which contains the Schwarzschild spacetime, ({\it 
b}) with $\Lambda > 0$, the Nariai \cite{Nar} solution and ({\it c}) 
with arbitrary $\Lambda$, the Kottler \cite{Kot} solution that 
includes the solutions in ({\it a}), de Sitter space (dS, $\Lambda > 
0$), anti-de Sitter space (AdS, $\Lambda < 0$) and Schwarzschild-dS 
(-AdS) spacetimes \cite{GH}.  The Nariai solution is important for its 
quantum instability: once created, it decays into a pair of near 
extreme black holes \cite{Bo}.  Also, generalizing the Kottler 
solution, there exist {\it topological} black holes asymptotic to anti-de 
Sitter space where the horizon can be a Riemann surface with arbitrary 
genus \cite{V}.  The topological black holes are relevant for testing 
(1) the so called ``AdS/CFT correspondence'' \cite{Mald} in higher 
spacetime dimensions and (2) the Horowitz-Polchinski \cite{HP} 
correspondence principle which provides microscopic interpretation of 
black hole entropy by linking a highly excited string state to a black 
hole (see also \cite{KV}).  It has been demonstrated \cite{SM} that 
topological black holes of arbitrary genus could result from the 
gravitational collapse of pressureless dust.  Thus the local geometry 
of {\sf STE} spacetimes with $\displaystyle{I_B}=0$ is of considerable 
interest.  In general, for the class of 4-dimensional {\sf STE} 
spacetimes, we find that the vanishing of $\displaystyle{I_B}$ implies 
that $^{(3)}\Sigma$ is locally foliated by 2-dimensional surfaces, 
$^{(2)}S$, of constant curvature.  From the resulting structure of 
$^{(3)}\Sigma$, we show that {\it any} nonflat static solution to 
$R_{\mu\nu} = \Lambda g_{\mu\nu}$ must be either Kottler-type or Nariai-type 
[subsection \ref{sub3.4}] {\it if} $\displaystyle{I_B}=0$.

\par  
To fix our notations, we recall that there is a coordinate system 
$\{x^{0},~x^{1},~x^{2},~x^{3}\}$ at each point of a static spacetime,
$(M^{4},~g)$, with $g_{0i}=0$ and $\partial_{0} g_{\mu\nu}=0$
for $i=1,~2,~3$ and $\mu,~\nu=0,1,~2,~3$. Here $\partial_{0}$ is the
timelike, twist-free Killing vector field on $(M^{4},~g)$. Thus,
$(M^{4},~g)$ may be locally represented as a warp-product \cite{On},
\begin{equation}
^{(3)}\Sigma \times_{f} \reals~~~ \mbox{with}\label{1.1}
\end{equation}
\begin{equation}
g = h_{ij}dx^{i}\otimes dx^{j} + f^{2} (- dx^{0}\otimes dx^{0})~,
\label{1.2}
\end{equation}
where $h_{ij} = g_{ij}$ is the Riemannian metric on the hypersurface 
$^{(3)}\Sigma~$, and the function $f$ on $^{(3)}\Sigma$ is defined by, 
$f = \sqrt {-g_{00}} > 0$. Throughout the paper, lower case Latin
indices will represent ``spatial'' coordinates on $^{(3)}\Sigma~$
and the Greek indices will denote spacetime coordinates. The curvature
tensor, Ricci 
tensor and scalar curvature on $^{(3)}\Sigma~$ will be denoted by 
${\hat R_{ijkl}}$, ${\hat R_{ij}}$ and ${\hat R}$, respectively.  On 
$^{(3)}\Sigma~$, we also define the Hessian of $f$ by the symmetric 
rank-2 tensor, $H(\partial_{i},~\partial_{j}) \equiv h\left ({\hat 
\nabla}_{\partial_{i}}{\widetilde {df}},~\partial_{j}\right ) \equiv 
H_{ij}$, where ${\widetilde {df}} \equiv (\partial^{k}f)\partial_{k} 
\equiv f^{k}\partial_{k}$ is the gradient of $f$, and ${\hat \nabla}$ 
is the Levi-Civita connection on ($~^{(3)}\Sigma,~h$).

\par The paper is organized as follows. In section \ref{sec2} 
we obtain an explicit formula for the invariant, $I \equiv 
R_{\mu\nu\rho\sigma;\delta}~R^{\mu\nu\rho\sigma;\delta}$, on a static 
spacetime and briefly outline its local measurability.  Here, the 
distinguished components, $\displaystyle{B_{ijk}\equiv R_{0ijk;0}}$, 
of the covariant derivative of curvature define a rank-3 tensor on 
$(~^{(3)}\Sigma,~h)$, and the scalar $\displaystyle{I_{B}\equiv 
R_{0ijk;0} R^{0ijk;0}}$ vanishes if and only if $R_{0ijk;0}=0$ since 
$h$ is Riemannian.  If a static spacetime is also an Einstein space, we 
prove in {\sf Proposition 1} that $\displaystyle{R_{0ijk;0} = (-f^{2}) 
{\hat R}_{ijk}}$, where ${\hat R}_{ijk}$ is the Cotton tensor \cite{E} 
on the Riemannian $3$-manifold, ($~^{(3)}\Sigma,~h$).

\par In section \ref{sec3} we consider a domain of 
($~^{(3)}\Sigma,~h$) where $||{\widetilde {df}}||^{2}\equiv 
h({\widetilde {df}},{\widetilde {df}}) \neq 0$ and determine the local 
geometry of {\sf STE} spacetimes subjected to the condition, 
$R_{0ijk;0}=0$.  First, we derive the characteristic structure of the 
Hessian of $f$, $H_{ij}$, which specifies the Weyl curvature operator 
for an {\sf STE} spacetime.  While any static spacetime is known to be 
either type $I$, $D$ or $O$, we prove that a static spacetime which is 
also an Einstein space must be locally of type $D$ or $O$ if 
$R_{0ijk;0} = 0$.  Furthermore, in the domain where $||{\widetilde 
{df}}||\neq 0$, the $1$-form $df$ defines a local foliation of 
($~^{(3)}\Sigma,~h$) and we determine the extrinsic and intrinsic 
geometry of the integral manifolds, $\displaystyle{^{(2)}S}$.  From 
the local structure of $~^{(3)}\Sigma$, we derive all non-trivial 
{\sf STE} spacetimes satisfying $R_{0ijk;0}=0$.  Finally, in section 
\ref{sec4} we discuss possible generalizations of our main results 
which can be summarized as follows.

\par\noindent {\sf Main Theorem:} Let $~^{(3)}\Sigma \times_{f} 
\reals$ be a static solution to $R_{\mu\nu} = \Lambda g_{\mu\nu}$ 
({\sf STE} spacetime).  (1) $~^{(3)}\Sigma$ is conformally flat if and 
only if $R_{0ijk;0} = 0$.  (2) The spacetime is locally of type $D$ or 
$O$ if $R_{0ijk;0} = 0$.  (3) $~^{(3)}\Sigma$ is locally foliated by 
totally umbilic 2-manifolds, $^{(2)}S$, of constant curvature if 
$R_{0ijk;0} = 0$ and hence, ($~^{(3)}\Sigma,~h$) is locally a 
warp-product space, $\reals\times$ $_{r(f)}$ $^{(2)}S$, for some 
function $r(f)$: if $r(f)$ is a constant, a non-trivial {\sf STE} 
spacetime is Nariai-type; otherwise it is Kottler-type.

\section{\sf Invariant, $I$, for static solutions  
($^{(3)}\Sigma \times_{f}\reals$) to
$R_{\mu\nu} = \Lambda g_{\mu\nu}$}\label{sec2}
\subsection{\sf Curvature components and explicit form of $I$}\label{sub2.1}

\par For a static spacetime, $M^{4}$ locally represented by $~ ^{(3)}\Sigma
\times_{f} \reals$, the nonzero covariant derivatives are
\begin{eqnarray}
	\nabla_{\partial_{i}}\partial_{j} & = & {\hat \Gamma}_{ij}^{k}
	{\partial_{k}},\label{2.1}\\
	\nabla_{\partial_{i}}\partial_{0} & = &
	\left(\frac{\partial_{i}f}{f}\right)\partial_{0}~,
~~\mbox{and}\label{2.2}\\
	\nabla_{\partial_{0}}\partial_{0} & = &
	f\left(\partial^{i}f\right)\partial_{i}~.\label{2.3}
\end{eqnarray}
\begin{remark}
	The functions,
	${\hat \Gamma}_{ij}^{k}$, in (\ref{2.1}) are the Christoffel components on
	$(~^{(3)}\Sigma,h)$. Thus, the second fundamental form of
	$^{(3)}\Sigma$ vanishes identically, and hence,
	$\nabla_{\partial_{i}}\partial_{j}
	= {\hat \nabla}_{\partial_{i}}\partial_{j}$, where
	${\hat \nabla}$ is the Levi-Civita connection on
    ($~^{(3)}\Sigma,~h$).
\end{remark}
Using (\ref{2.1}) - (\ref{2.3}), components of the curvature tensor 
for any static spacetime $M^{4}$ can be expressed in terms of the data 
on $^{(3)}\Sigma$:
\begin{eqnarray}
	R_{0ijk} & = & 0~,\label{2.4} \\
	R_{0i0j} & = & f H_{ij}~,~~\mbox{and}\label{2.5}\\
	R_{ijkl} & = & {\hat R}_{ijkl}~,\label{2.6}
\end{eqnarray}
where the symmetric tensor field, $H_{ij}\equiv
H(\partial_{i},~\partial_{j})$ represents
the Hessian of the function $f(x^{1},~x^{2},~x^{3})$ on
($~^{(3)}\Sigma,~h$),
\begin{equation}
	H(\partial_{i},~\partial_{j}) =
	\partial_{i}\partial_{j}f - {\hat \Gamma}^{k}_{ij}
	\partial_{k}f.\label{2.7}
\end{equation}
\begin{remark}
	Since the extrinsic curvature of $(^{(3)}\Sigma,~h)$ is zero, the
	``spatial'' components of the curvature tensor for the spacetime
	$(M^{4},~g)$ are given by $R_{ijkl} = {\hat R}_{ijkl}~$, where 
	${\hat R}_{ijkl}$ is described below.
\end{remark}
\noindent
The curvature tensor, ${\hat R}_{ijkl}$, of $(^{(3)}\Sigma,~h)~$, is
given by
\begin{eqnarray}
	{\hat R}_{ijkl} & = & \frac{{\hat R}}{2}
	\left \{h_{il}h_{jk} - h_{ik}h_{jl}\right \}\nonumber\\
	& + & \left \{{\hat R}_{ik}h_{jl} - {\hat R}_{il}h_{jk}\right \}
	+ \left \{{\hat R}_{jl}h_{ik} - {\hat R}_{jk}h_{il}\right \}~,
	\label{2.8}
\end{eqnarray}
where the Ricci tensor and scalar curvature of $(^{(3)}\Sigma,~h)$ are
related to the Ricci tensor, $R_{\mu\nu}$, of $(M^{4},~g)$ by
\begin{eqnarray}
	{\hat R}_{ij} & = & \frac{1}{f} H_{ij} + R_{ij}~~~\mbox{and }
	\label{2.9}\\
	{\hat R} & \equiv & h^{ij}{\hat R}_{ij} = \frac{R_{00}}{f^{2}} +
	h^{ij}R_{ij}~.\label{2.10}
\end{eqnarray}
For a static spacetime, $(M^{4},~g)$, the scalar invariant, 
$I\equiv R_{\mu\nu\rho\sigma;\delta}
R^{\mu\nu\rho\sigma;\delta}$, is given by
\begin{equation}
	I = h^{mn} R_{ijkl;m}
	{R^{ijkl}}_{;n}
	+ 4 h^{mn} R_{0i0j;m}{R^{0i0j}}_{;n}
	+ 4 (g^{00})^{2} R_{0ijk;0}{{R_{0}}^{ijk}}_{;0}\label{2.11}
\end{equation}
The quantities appearing in the above equation are derived from 
locally measurable components of the covariant 
derivative of the curvature tensor \cite{K2}.  Using normal 
coordinates $\xi^{A}$ in a field of frames parallel 
transported along the geodesics from $\xi^{A} = 0$, the curvature 
tensor can be expanded as \cite{W}
\[
	R_{ABCD} (\xi) = R_{ABCD} (0) + R_{ABCD;E} (0) \xi^{E} + 
	O(\xi^{2}).
\]
From the equation for geodesic deviation, one finds the curvature 
components by measuring \cite{MTW} relative accelerations of test 
particles moving with various speed and directions in the neighborhood 
of $\xi^{A} = 0$.  Hence, the covariant derivatives of the curvature 
tensor at $\xi^{A} = 0$ can be obtained from the above expansion.

\par \noindent 
If a static spacetime, $(M^{4},g)$, is also an Einstein space,
\begin{equation}
	R_{\mu\nu} = \Lambda g_{\mu\nu},\label{3.1}
\end{equation}
for some constant $\Lambda\in\reals$, then we have 
$R_{ij} = \Lambda h_{ij}$, $R_{0j}
= 0$ and $R_{00} = -\Lambda f^{2}$.  Also, from the equations
(\ref{2.9}) and (\ref{2.10}), the Ricci
tensor and scalar curvature on $(^{(3)}\Sigma,~h)$ are given by 
\begin{eqnarray}
	{\hat R}_{ij} & = & \frac{1}{f} H_{ij} + \Lambda h_{ij},
	\label{3.2}\\
	{\hat R} & = & h^{ij}{\hat R}_{ij} = \frac{1}{f} (tr H) +
	h^{ij}R_{ij} = 2\Lambda,\label{3.3}
\end{eqnarray}
where $\frac{1}{f} (tr H) = -\Lambda$.  Inserting (\ref{2.4})-(\ref{2.8}) 
and (\ref{3.2})-(\ref{3.3}) in (\ref{2.11}) we have
\begin{equation}
I = 8~{\hat R}_{ij;k} {\hat R}^{ij;k} + {\frac {4}
{f^{4}}} R_{0ijk;0}{{R_{0}}^{ijk}}_{;0}.\label{3.4}
\end{equation}

\subsection{\sf Probing conformal flatness of $^{(3)}\Sigma$}
\label{sub2.2}
\par As noted in section \ref{sec1}, the significance of the second term
in the scalar invariant, $I$, for the
local geometry of an {\sf STE} spacetime is derived from the
relation between $R_{0ijk;0}$ and the Cotton tensor. This relation is
established in the following proposition.

\noindent {\sf Proposition 1}: In a static spacetime ($^{(3)}\Sigma 
\times_{f} \reals$), $R_{0ijk;0}$ defines a rank-3 tensor field on 
$(^{(3)}\Sigma,~h)$.  If ($^{(3)}\Sigma \times_{f} \reals$) is also an 
Einstein space then $R_{0ijk;0}$ is related to the Cotton tensor, 
${\hat R}_{ijk}$, by
\begin{equation}
	R_{0ijk;0} = (-f^{2}) {\hat R}_{ijk}.\label{3.5}
\end{equation}
\noindent {\sl Proof.}
The contracted Bianchi identity \cite{HE},
$R^{\mu}_{\nu\rho\sigma;\mu} = R_{\nu\sigma;\rho} - R_{\nu\rho;\sigma}$,
implies that
\begin{eqnarray}
	R^{0}_{ijk;0} + R^{m}_{ijk;m} & = & R_{ik;j} - R_{ij;k}.\nonumber
\end{eqnarray}
Using (\ref{2.6}) and (\ref{2.9}) in the above equation, and inserting 
(\ref{3.2})-(\ref{3.3}) in the definition of the 
Cotton tensor, 
${\hat R}_{ijk} = ({\hat R}_{ij;k} - {\hat R}_{ik;j})
           -{\frac {1}{4}} (h_{ij}{\hat R}_{;k} - h_{ik}{\hat
           R}_{;j})$,
we have
\begin{equation}
\left (-\frac{1}{f^{2}}\right)R_{0ijk;0} = \left (\frac{1}{f} H_{ij}\right
)_{;k}
- \left (\frac{1}{f} H_{ik}\right )_{;j} = {\hat R}_{ijk}.\label{3.6}
\end{equation}
\noindent Now, {\sf Proposition 1} follows from the equation 
(\ref{3.6}), where the first equality holds for any static spacetime.

\par
Using (\ref{3.5}) in  (\ref{3.4}), the scalar invariant for 
an {\sf STE} spacetime is now given by
\begin{equation}
I = 8~{\hat R}_{ij;k} {\hat R}^{ij;k} +
4~ {\hat R}_{ijk}{\hat R}^{ijk}~.\label{3.7}
\end{equation}
\par\noindent Thus, in $I$ [(\ref{3.4})],  
vanishing of the  
term, $\displaystyle{I_{B}\equiv R_{0ijk;0} R^{0ijk;0} 
= {\hat R}_{ijk}{\hat R}^{ijk}}$, is  
equivalent to conformal flatness of $(~^{(3)}\Sigma,~h)$ 
since a 3-dimensional Riemannian manifold is conformally flat if and 
only if the Cotton tensor, ${\hat R}_{ijk} = 0$ \cite{E}.  This proves 
the part-(1) of the {\sf Main Theorem} as stated at the end of  
section \ref{sec1}.

\par For the Schwarzschild spacetime with $\Lambda = 0$, $f^{2} = (1 - 
r_{s}/r)$ and $h = (dr\otimes dr)/f^{2} + r^{2} (d\theta\otimes 
d\theta + sin^{2}\theta~~d\phi\otimes d\phi)$, the term 
$\displaystyle{I_{B}\equiv R_{0ijk;0} R^{0ijk;0}}$ in $I$ 
vanishes, and hence,
\[
I = 8 {\hat R}_{ij;m} {\hat R}^{ij;m} = 180 {\frac
{r_{s}^{2}}{r^{8}}} \left
(1 - {\frac {r_{s}}{r}}\right ),\label{1.3}
\]
where $r_{s}$ the Schwarzschild radius.  Hence, the local 
measurement of the scalar invariant, $I$, not only 
reveals the passage through the event horizon \cite{K2} but also the 
conformal flatness of the spacelike hypersurface, ($~^{(3)}\Sigma,~h$).
In the next section we turn to further consequences of
$\displaystyle{I_{B}\equiv R_{0ijk;0} R^{0ijk;0}}=0$ for the 
local geometry of {\sf STE} spacetimes. 

\section{\sf Conformally flat $^{(3)}\Sigma$ and static solutions to
$R_{\mu\nu} = \Lambda g_{\mu\nu}$}\label{sec3}

\par To prove the part-(2) and part-(3) of the {\sf Main Theorem}, 
we obtain
an explicit form of $R_{0ijk;0}$ for an {\sf STE} spacetime.
For a static spacetime, it
follows from the first equality in (\ref{3.6}) that
\begin{eqnarray}
R_{0ijk;0} 
& = & f_{k} H_{ij} -f_{j} H_{ik} - f (H_{ij;~k} - H_{ik;~j}),\label{3.8}
\end{eqnarray}
where $f_{i}\equiv \partial_{i}f$. Since $H_{ij}
=\partial_{i}\partial_{j}f - {\hat \Gamma}^{k}_{ij} \partial_{k}f = f_{i;j}$,
the Ricci identity for the
covariant vector field, $f_{i}$, on $(^{(3)}\Sigma,~h)$, gives
\begin{eqnarray}
H_{ij;k} - H_{ik;j} & = & f_{i;jk} - f_{i;kj} = f^{m} {\hat 
R}_{mijk}.\nonumber
\end{eqnarray}
Inserting the above identity in (\ref{3.8}), we find
\begin{eqnarray}
R_{0ijk;0} & = &  H_{ij}f_{k}
                - H_{ik}f_{j} - f f^{m} {\hat R}_{mijk}.\label{3.9}
\end{eqnarray}
\par\noindent Replacing
${\hat R}_{mijk}$ in (\ref{3.9}) by (\ref{2.8}) and using 
(\ref{3.2})-(\ref{3.3}) we have
\begin{equation}
	R_{0ijk;0} = 2 (H_{ij} f_{k} - H_{ik} f_{j})
	- h_{ij} \{ (tr H) f_{k} - H_{km} f^{m}\}
	+ h_{ik} \{ (tr H) f_{j} - H_{jm} f^{m}\},\label{3.10}
\end{equation}
in {\sf STE} spacetimes. Here and in the rest of the paper, 
we consider a domain of ($~^{(3)}\Sigma,~h$) where $||{\widetilde 
{df}}||^{2}\equiv h({\widetilde {df}},{\widetilde {df}}) \neq 0$ and 
hence, the 1-form $df$ defines a local foliation of $~^{(3)}\Sigma$.  
Integral manifolds of this foliation are spacelike hypersurfaces, 
$^{(2)}S$, which are, locally, $f=constant$ surfaces in 
($~^{(3)}\Sigma,~h$).  The local vector field $\displaystyle{N_{k} 
\equiv f_{k}/||{\widetilde {df}}||}$ is the unit normal, and the 
symmetric tensor field, ${\bar h}_{ij} = h_{ij} - N_{i}N_{j}$, is the 
induced metric on $^{(2)}S \subset ~^{(3)}\Sigma$. 

\subsection{\sf Structure of the Hessian of $f$}\label{sub3.1}

\par First, we prove a basic result that $R_{0ijk;0} = 0$ 
(or conformal flatness of $~^{(3)}\Sigma$) imprints a 
characteristic structure on the Hessian $H_{ij}$ in {\sf STE} spacetimes: 
$H_{ij}$ has an eigenvalue $H(N,N)$ with the 1-dimensional 
eigenspace along $N$, and a repeated eigenvalue $\{(tr H) - 
H(N,N)\}/2$ with the tangent plane to $^{(2)}S$ as the 2-dimensional 
eigenspace, at each point of the regular domain of $^{(3)}\Sigma$.

\par\noindent {\sf Proposition 2}: In a domain of an {\sf STE}
spacetime ($^{(3)}\Sigma \times_{f} \reals$) where $df\neq 0$, the
rank-3 tensor, $R_{0ijk;0}$, on ($~^{(3)}\Sigma,~h$) vanishes {\it  if and
only if}
\begin{equation}
	H_{ij} = \frac{1}{2}\{(tr H) - H(N,N)\}{\bar h}_{ij}
	+ H(N,N)N_{i}N_{j},\label{3.11}
\end{equation}
where $N = {\widetilde {df}}/||{\widetilde {df}}||$ is the unit normal
field on the hypersuface, 
$(~^{(2)}S,~{\bar h})\subset (~^{(3)}\Sigma,~h)$.

\noindent{\sl Proof.}
Setting $R_{0ijk;0} = 0$ and $f_{k} = (||{\widetilde {df}}||)N_{k}$ in 
(\ref{3.10}), we have
\begin{equation}
	2 (H_{ij} N_{k} - H_{ik} N_{j})
	- h_{ij} \{ (tr H) N_{k} - H_{km} N^{m}\}
	+ h_{ik} \{ (tr H) N_{j} - H_{jm} N^{m}\} = 0. \label{3.12}
\end{equation}
Contracting both sides of
(\ref{3.12}) by $N^{i}N^{k}$, we find
\begin{equation}
H_{ij}N^{i} = H(N,N)N_{j}.\label{3.13}
\end{equation}
Now, inserting (\ref{3.13}) back in (\ref{3.12}), leads to
\begin{equation}
	\{2 H_{ij} - (tr H - H(N,N))h_{ij}\}N_{k}
   -\{2 H_{ik} - (tr H - H(N,N))h_{ik}\}N_{j}
   = 0.\label{3.14}
\end{equation}
Finally, contracting (\ref{3.14}) by $N^{k}$, we have (\ref{3.11}).
Conversely, if the equation (\ref{3.11}) holds, then contracting with
$N^{i}$, we get (\ref{3.13}).  Now, inserting (\ref{3.11}) and
(\ref{3.13}) in (\ref{3.10}), we find $R_{0ijk;0} = 0$.

\subsection{\sf Petrov types}

\par\noindent {\sf Corollary}: For an {\sf STE} spacetime,
$R_{0ijk;0} = 0$ implies that the spacetime is locally of 
type $D$ or $O$.
\par\noindent{\sl Proof.}
We choose an oriented orthonormal basis $\{e_{0}=\partial_{0}/f,~ e_{1} 
= N,~e_{2},~e_{3}\}$ for the tangent space at each point of the spacetime. 
Using this basis, the matrix of the Weyl curvature operator  
- a self adjoint linear transformation 
\begin{equation}
	{\bf C}:\Lambda^{2}_{*} \longrightarrow \Lambda^{2}_{*}
\end{equation}
on the 3-dimensional complex vector space of bivectors 
($\Lambda^{2}_{*}$) - is given by \cite{On2}
\begin{equation}
	[{\bf C}] = -A + iB,\label{matC}
\end{equation}
where $3\times 3$ real symmetric traceless matrices $A$ and $B$ have 
the following form:
\begin{eqnarray}
	A_{0i,0j} & = & C (e_{0}\wedge e_{i},~e_{0}\wedge e_{j}) \equiv 
	C (e_{0},~e_{i},~e_{0},~e_{j}),\label{matA} \\
	B_{0i,jk} & = & C (e_{0}\wedge e_{i},~e_{j}\wedge e_{k}) \equiv 
	C (e_{0},~e_{i},~e_{j},~e_{k}),\label{matB}
\end{eqnarray}
for $i,~j,~k = 1,2,3$.  Here, a double index ``$\mu\nu$'' represents a 
basis element, $e_{\mu}\wedge e_{\nu}$, of the real bivector space, 
$\Lambda^{2}$; the Weyl cuvature tensor, $C$, induces a symmetric 
bilinear form (also denoted by $C$) on $\Lambda^{2}$, 
\begin{equation}
	C:\Lambda^{2}\times \Lambda^{2} \longrightarrow \reals,\nonumber 
\end{equation}
defined by $C (e_{\mu}\wedge e_{\nu},~e_{\rho}\wedge e_{\sigma}) 
\equiv C (e_{\mu},~e_{\nu},~e_{\rho},~e_{\sigma})$; $A_{0i,0j}$ and 
$B_{0i,jk}$ are the distinguished matrix elements of this bilinear 
form (see \cite{On2} for the details). 
For an {\sf STE} spacetime, the equations
(\ref{matA}) and (\ref{matB}), give
\begin{eqnarray}
	A_{0i,0j} & = & \frac{1}{f}\left\{H(e_{i}, e_{j})
	- \left (\frac{tr H}{3}\right)g(e_i,e_j)\right\},\label{vA} \\
	B_{0i,jk} & = & 0.\label{vB}
\end{eqnarray}
If $R_{0ijk;0} = 0$, it follows from the equation (\ref{3.11}) in 
{\sf Proposition 2} that
\begin{eqnarray}
H(e_{i}, e_{j}) & = & 0,~~~\mbox{for $i\neq j$}\label{offDg} \\
 & = & \frac{1}{2}\left\{tr H - H(e_{1}, e_{1})\right\}~~~\mbox{for
 $i=j=2,~3$}.\label{compH}
\end{eqnarray}
Now, the matrix of the Weyl curvature operator can be obtained
from the equations (\ref{matC}) and (\ref{vA})-(\ref{compH}),
\[
[{\bf C}] =\left(
\begin{array}{ccc}
	 \lambda_{1} & 0 & 0  \\
	0 &  \lambda_{2} & 0  \\
	0 & 0 & \lambda_{3}
\end{array} \right),
\]
\noindent where $\displaystyle{\lambda_{2} = \lambda_{3} = 
-\lambda_{1}/2}$ and $\displaystyle{\lambda_{1} = -\left\{H(N,N) - 
(tr H)/3\right\}/f}$. Hence, the spacetime is locally of type $D$ if 
$\lambda_{1}\neq 0$, and type $O$ if $\lambda_{1} = 0$. This proves 
part-(2) of the {\sf Main Theorem}.
\begin{remark}
\par Our solution for $H_{ij}$ in (\ref{3.11}) [{\sf Proposition 
2}] can be written as
\[
H_{ij} = \frac{1}{2}\{(tr H) - H(N,N)\} h_{ij}
	+ \frac{3}{2}\{H(N,N) - (tr H)/3\}N_{i}N_{j},
\]
where we used ${\bar h}_{ij} = h_{ij} - N_{i}N_{j}$. Hence,
the solution, $H_{ij} = \{(tr H) - H(N,N)\} h_{ij}/2$,
is {\it a special case} of our solution in (\ref{3.11}) when 
$\{H(N,N) - (tr H)/3\}=0$. In fact, this special case corresponds to
$\lambda_{1} = 0$ that leads to a locally type $O$ spacetime.
\end{remark}

\subsection{\sf Local warp-product structure of  
$^{(3)}\Sigma$}\label{sub3.3}

\par To complete the proof of the {\sf Main Theorem}, we need to
characterise the extrinsic geometry of $^{(2)}S \subset
~^{(3)}\Sigma$ in {\sf STE} spacetimes when $R_{0ijk;0} = 0$.
Using the projection tensor,
${\bar h}^{i}_{j} ={\delta}^{i}_{j} - N^{i}N_{j}$, we introduce the
extrinsic curvature \cite{HE} of $^{(2)}S$,
\begin{equation}
	K_{ij}\equiv {\bar h}^{m}_{i}{\bar h}^{k}_{j}N_{k;m}
	=\frac{1}{||{\widetilde {df}}||}{\bar h}^{m}_{i}{\bar h}^{k}_{j}
	H_{mk}.\label{ecur}
\end{equation}
Now, we observe that
the equation (\ref{3.11}) holds if and only if
\begin{eqnarray}
		H_{il} N^{i}{\bar h}^{l}_{m} & = & 0 ~~~\mbox{and}\nonumber\\
		{\bar h}^{m}_{i}{\bar h}^{k}_{j} H_{mk} & = &
		\frac{1}{2}\{tr H - H(N,N)\}{\bar h}_{ij}.\label{eqval}
\end{eqnarray}
Then, from (\ref{ecur})-(\ref{eqval}) and {\sf Propostion 2}, 
we have $R_{0ijk;0} = 0$ if and only if
\begin{equation}
	H_{il} N^{i}{\bar h}^{l}_{m}  =  0 ~~~\mbox{and}~~~
	K_{ij} = \left (\frac{tr K}{2}\right) {\bar h}_{ij},\label{Kij}
\end{equation}
where $\displaystyle{tr K \equiv h^{ij}K_{ij} = 
\left\{tr H - H_{ij} N^{i}N^{j}\right\}/{||{\widetilde {df}}||}}$. 
The {\it first equation} in (\ref{Kij}) implies that $||{\widetilde {df}}||$ is 
a function of $f$: this follows from the equality,
$\displaystyle{H_{il} N^{i}{\bar h}^{l}_{m}
= (1/2||{\widetilde {df}}||)\left(||{\widetilde 
{df}}||^{2}\right)_{||m}}$, where $`||'$ denotes covariant derivative on 
$(~^{(2)}S,~{\bar h})$;  this equality can be derived from
\begin{eqnarray}
	\left(h^{ij} f_{i} f_{j}\right)_{||m}
	& = & 2 H_{il} f^{i}{\bar h}^{l}_{m}~.\label{covd}
\end{eqnarray} 
The {\it second equation} in (\ref{Kij}) shows that the
hypersurface $^{(2)}S \subset ~^{(3)}\Sigma$ in {\sf STE} spacetimes 
is totally umbilic
(and hence, locally orientable) if $R_{0ijk;0} = 0$. Next, we show that
the mean curvature, (${tr K}$), depends only on $f$.
\par\noindent {\sf Lemma 1:} The mean curvature, $({tr K})$, 
of the the hypersurface, $(~^{(2)}S,~{\bar h})\subset 
(~^{(3)}\Sigma,~h)$ in {\sf STE} spacetimes is constant on 
$~^{(2)}S$ if $R_{0ijk;0} = 0$.

\noindent{\sl Proof.}  From (\ref{Kij}), we insert $K^{l}_{i} = \frac{1}{2}
(tr K) {\bar h}^{l}_{i}$ in the Codazzi equation \cite{HE} for the
hypersurface, $(~^{(2)}S,~{\bar h})\subset (~^{(3)}\Sigma,~h)$:
\begin{eqnarray}
	{\hat R}_{mj} N^{j} {\bar h}^{m}_{i}
	& = & K^{l}_{i||l} - (tr K)_{||i}
	= -\frac{1}{2}(tr K)_{||i}~.\label{scurva}
\end{eqnarray}
From (\ref{3.2}) and (\ref{eqval}), we also have
\begin{equation}
	{\hat R}_{mj} N^{j} {\bar h}^{m}_{i}
	=(\frac{1}{f} H_{mj} + \Lambda h_{mj})N^{j} {\bar h}^{m}_{i} =
0~.\label{scurvb}
\end{equation}
From (\ref{scurva})-(\ref{scurvb}), it follows that $(tr K)_{||i} = 0$.

\par We recall that the scalar
curvature, ${\bar R}$, of the $f = constant$ hypersurface, $(~^{(2)}S,~{\bar
h})\subset (~^{(3)}\Sigma,~h)$, is given by the Gauss equation
\cite{HE},
\begin{equation}
	{\bar R}  =  {\hat R} - 2{\hat R}_{ij} N^{i}N^{j} +
	(tr K)^{2} - tr K^{2}.\label{(3.24)}
\end{equation}
Using (\ref{Kij}) as well as the relations ${\hat R} =2\Lambda$ and
${\hat R}_{ij} N^{i}N^{j} =\frac{1}{f} H_{ij} N^{i}N^{j} + \Lambda$
- derived from the equations (\ref{3.2})-(\ref{3.3}), we find
\begin{equation}
	{\bar R} = \frac{2}{f}\left\{||{\widetilde {df}}||(tr K) + 
	\Lambda f\right\}
	+\frac{1}{2}(tr K)^{2}. \label{(3.25)} 
\end{equation}
Since $||{\widetilde {df}}||$ is a function of $f$ by 
(\ref{Kij})-(\ref{covd}), it follows from the {\sf Lemma 1} 
that ${\bar R}(f)$ is constant on the level surfaces, $^{(2)}S$, of $f$.  
Thus, in a {\sf 
STE} spacetime if $R_{0ijk;0} = 0$ or equivalently, $(~^{(3)}\Sigma, 
h)$ is conformally flat, then $~^{(3)}\Sigma$ is locally foliated by 
totally umbilic [(\ref{Kij})] surfaces [$^{(2)}S$] of constant 
curvature [(\ref{(3.25)})], and hence, $(~^{(3)}\Sigma, h)$ is locally 
a warp-product space, $\reals\times$$_{r(f)}$$^{(2)}S$, for 
some function $r(f)$:
\begin{equation}
h = \frac{1}{||{\widetilde {df}}||^{2}} df\otimes df + 
r(f)^{2} \sigma_{k},\label{3.26}
\end{equation}
where $\sigma_{k}$ is the metric of constant curvature 
$k=0,\pm 1$ on $^{(2)}S$. 

\subsection{\sf Non-trivial static solutions to 
$R_{\mu\nu} = \Lambda g_{\mu\nu}$}
\label{sub3.4}

If $(~^{(3)}\Sigma, h)$ is conformally flat, then the general form of 
the static solutions to $R_{\mu\nu} = \Lambda g_{\mu\nu}$ is given by
\begin{equation}
	g = -f^{2}dx^{0}\otimes dx^{0} + u(f)^{2} df\otimes df + 
	r(f)^{2} \sigma_{k},\label{gste}
\end{equation}
where we have used (\ref{1.2}) as well as (\ref{3.26}), and set 
$u(f)\equiv 1/||{\widetilde {df}}||$. The functions $r(f)$ and $u(f)$ 
must satisfy the field equations [(\ref{3.2})-(\ref{3.3})] 
\begin{eqnarray}
\frac{r''}{r} - \frac{r'}{r}\frac{u'}{u} & = & 
-\frac{1}{2}\Lambda u^{2} + \frac{1}{2f}\frac{u'}{u},\label{3.27}\\
2\frac{r'}{r} - \frac{u'}{u} & = & -\Lambda f u^{2},\label{3.28}
\end{eqnarray}
where a {\it prime} denotes differentiation with respect to $f$.

\par Assuming $r(f)$ is invertible, we set 
$\displaystyle{\psi (r) = (df/dr) u(f(r))}$ in (\ref{gste}), and 
the solution to (\ref{3.27})-(\ref{3.28}) is given by 
$\psi (r) = 1/f(r)$ where 
$f^{2} (r) = k - 2\mu/r - (\Lambda/3)r^{2}$. This leads to 
spacetimes in Kottler class,
\begin{equation}
g_{kot} =-\left (k - \frac{2\mu}{r} - \frac{\Lambda}{3} 
r^{2}\right) dx^{0}\otimes dx^{0} + 
\left (k - \frac{2\mu}{r} - \frac{\Lambda}{3} 
r^{2}\right)^{-1} dr\otimes dr + r^{2}\sigma_{k},~~ 
\mu\in\reals,\label{3.29} 
\end{equation}

\par If $r(f) = c$ is a {\it nonzero constant}, it follows from (\ref{gste}) 
and $R_{\mu\nu} = \Lambda g_{\mu\nu}$ that $k/c^{2} = \Lambda$. For 
$k\neq 0$ (or equivalently, $\Lambda\neq 0$), 
one obtains spacetimes in Nariai class \cite{Nar}),
\begin{equation} 
g_{nar} =-\left (a - \Lambda\rho^{2}\right) dx^{0}\otimes dx^{0} +
\left (a - \Lambda\rho^{2}\right)^{-1} d\rho\otimes d\rho + 
\frac{1}{|\Lambda |} \sigma_{k},~~k=\frac{\Lambda}{|\Lambda |},~~
a\in\reals,\label{3.30}
\end{equation}
where $\rho \equiv ||{\widetilde {df}}||/|\Lambda |$ and 
$f^{2} = a - \Lambda\rho^{2}$.

\par Finally, we consider two remaining possible special cases of 
(\ref{gste}).  (i) $r(f) = c$ and $k=0$ which yields $\Lambda = 0$, 
and (ii) $u(f)$ is a constant which implies $r(f)$ is a constant and 
hence, $\Lambda = 0$.  Each of these cases leads to a flat metric.  
Thus, conformal flatness of $(~^{(3)}\Sigma, h)$ implies that any 
non-trivial static solution solution to $R_{\mu\nu} = \Lambda 
g_{\mu\nu}$ must be of Kottler-type [(\ref{3.29})] or Nariai-type 
[(\ref{3.30})].  This completes the proof of the {\sf Main Theorem}.

\section{\sf Conclusions}\label{sec4}
\par\noindent We conclude with some remarks on possible
generalizations of the {\sf Main Theorem}. 
\par\noindent
(a) The relation, $R_{0ijk;0}
= (-f^{2}) {\hat R}_{ijk}$, in (\ref{3.5}) holds for an $n$-dimensional
static spacetime, $~^{(n-1)}\Sigma \times_{f} \reals$, which is also an 
Einstein space. Here,
the Cotton tensor of $~^{(n-1)}\Sigma$ is given by
\[
{\hat R}_{ijk} = ({\hat R}_{ij;k} - {\hat R}_{ik;j})
           -{\frac {1}{2(n-2)}} (h_{ij}{\hat R}_{;k} - h_{ik}{\hat
           R}_{;j}).\nonumber
\]
\par\noindent
(b) For $n>4$, the Cotton tensor of $~^{(n-1)}\Sigma$
vanishes if and only if its Weyl curvature is
divergence free \cite{E}:
\[
{{\hat C}^{m}~}_{ijk;m} = -\frac{(n-4)}{(n-3)} {\hat R}_{ijk}.
\]
Thus, the conclusion of part-(3) of the {\sf Main Theorem} holds 
for an $n$-dimensional
{\sf STE} spacetime if $~^{(n-1)}\Sigma$ is conformally flat for
$n > 4$.
An example of this case is the
$Schwarzschild-AdS_{5}$ static solution (see Appendix of \cite{CHR})
to the 5-dimensional Einstein equations with a cosmological constant.
The recent proof of the {\it Riemannian Penrose inequality} 
\cite{HuI} has established the lower bound for the mass of an 
asymptotically flat black hole spacetime. It would be 
interesting to see if and how this rigorous result can be extended to 
$n$-dimensional ($n\geq 4$) toplogical black holes which are
asymptotically locally AdS.
\par\noindent
(c) For a stationary spacetime $(M^{4},~g)$ there is a coordinate
system $\{x^{0},~x^{1},~x^{2},~x^{3}\}$ at each point of $M^{4}$ with
$\partial_{0}
g_{\mu\nu}=0~$, and
\[
g = - f^{2} (dx^{0} + \omega)\otimes (dx^{0} + \omega) +
h_{ij}dx^{i}\otimes dx^{j}, \nonumber
\]
where $g_{00} = - f^{2}$, $\omega \equiv\omega_{i} dx^{i}$, 
$\omega_{i} \equiv {g_{0i}}/{g_{00}}$, and $h_{ij} \equiv g_{ij} - 
g_{00} \omega_{i} \omega_{j}$, for $i,~j =1,~2,~3$ and 
$\mu,~\nu=0,1,~2,~3$.  The 2-tensor, $h_{ij}$, is the metric on the 
space of integral curves \cite{Ha} of the timelike Killing vector field 
($\partial_{0}$), and $R_{0ijk;0}$ defines a rank-3 tensor on this 
space. In this case, the structure of $R_{0ijk;0}$ is more
complicated although the equation (\ref{3.5}) holds in the static limit
($\omega = 0$).  For the Kerr spacetime \cite{HE}, one can check that
$R_{0ijk;0}$ vanishes.  Thus, it may be worthwhile to
examine the implications of $R_{0ijk;0} = 0$ for the local geometry of
stationary spacetimes which are Einstein spaces.

\acknowledgments  
One of us (MM) would like to thank Joel Hass and Steven Carlip for helpful discussion. This work is supported in part by the U.S. Department of Energy under 
Grant No DE-FG02-84ER40153.

\end{document}